\documentstyle[12pt,oldlfont]{article}

\begin{document}

\newcommand{\non}{\nonumber}
\newcommand{\nn}{\noindent}
\newcommand{\s}{\\ \vspace*{-3mm}}
\newcommand{\nt}{\not \hspace*{-1mm} }
\newcommand{\be}{\begin{eqnarray}}
\newcommand{\en}{\end{eqnarray}}
\newcommand{\lam}{\lambda^{\frac{1}{2}}}
\newcommand{\ra}{\rightarrow}

\renewcommand{\thefootnote}{\fnsymbol{footnote} }

\nn \hspace*{11.5cm} UdeM-GPP-TH-94-02 \\
    \hspace*{11.5cm} NYU--TH--94/06/01 \\
\hspace*{11.5cm} June 1994

\vspace*{1.2cm}

\centerline{\large{\bf QCD corrections to Higgs boson}}

\vspace*{0.4cm}

\centerline{\large{\bf self--energies and fermionic decay widths.} }

\vspace*{1.5cm}

\centerline{\sc A.~Djouadi$^1$ and P.~Gambino$^2$.}

\vspace*{1cm}

\centerline{$^1$ Groupe de Physique des Particules, Universit\'e de
Montr\'eal,  Case 6128 Suc.~A,}
\centerline{H3C 3J7 Montr\'eal PQ, Canada.}

\vspace*{0.4cm}

\centerline{$^2$ Department of Physics, New York University, 4 Washington
Place,}
\centerline{ New York, NY 10003, USA.}

\vspace*{2cm}

\begin{center}
\parbox{14cm}
{\begin{center} ABSTRACT \end{center}
\vspace*{0.2cm}

\nn We present the QCD corrections to Higgs boson self--energies for arbitrary
momentum transfer and for different internal quark masses, to treat the case of
CP--even, CP--odd and charged Higgs bosons which appear in extensions of the
Standard Model scalar sector. Using Ward Identities, we then relate these
results obtained by directly evaluating the relevant two--loop Feynman diagrams
to the known expressions for the electroweak vector boson vacuum polarization
functions. Finally, we derive the exact analytical expressions for the QCD
corrections to the decays of these Higgs particles into quark pairs in the
general case, and reproduce in a completely independent way known results in
some special cases. }

\end{center}

\newpage

\renewcommand{\thefootnote}{\arabic{footnote} }
\setcounter{footnote}{0}

\subsection*{1.~Introduction}

\renewcommand{\theequation}{1.\arabic{equation}}
\setcounter{equation}{0}

One of the standing enigmas in the modern formulation of the theory of strong
and electroweak interactions is the mechanism of electroweak symmetry breaking.
The existence of at least one scalar particle, the Higgs boson, is required to
generate the masses of the other fundamental particles, leptons, quarks and
weak gauge bosons \cite{HHG}. The discovery of this particle and the study of
its fundamental properties will be the most important mission of future
high--energy colliders \cite{REV}. \s

The phenomenological properties of the unique Higgs particle of the Standard
Model (SM) have been studied in great details in the literature \cite{HHG,REV}.
In fact, because the precise knowledge of Higgs decay widths, branching
fractions and production cross sections is mandatory, quantum corrections must
be included and this subject has received much attention recently \cite{KN}. In
particular, QCD corrections to Higgs decay and production processes are of the
utmost importance. For instance, the Higgs decays into quark pairs and gluons,
which together with the $H\rightarrow \tau^+ \tau^-$ decays, are the most
important decay modes in the intermediate mass range, $M_W < M_H < 140$ GeV,
receive very large QCD corrections. In the case of $H \rightarrow  q\bar{q}$,
they are known exactly to ${\cal O}(\alpha_S)$ \cite{RS,RA,RRY} and up to
${\cal
O}(\alpha_S^2)$ \cite{RS2} in the approximation $m_q \ll M_H$; in the case of
the gluonic decays, the QCD corrections are known up to
next--to--leading--order
\cite{Hgg}.\s

The electroweak radiative corrections to Higgs decays \cite{EW} are also of
significance, since the leading contribution is quadratically proportional to
the mass of the heavy top quark \cite{CDF}. In fact, a fourth generation of
heavy fermions, the existence of which is still allowed by present experimental
data with the proviso that the associated neutrino is heavy enough \cite{LEP},
would have a dramatic effect on the Higgs decay widths. Its contribution is
universal in the sense that it does not depend on the final state particle, and
will also increase quadratically with the heavy fermion masses. The universal
part of the two--loop mixed ${\cal O}(\alpha_S G_F m_Q^2$) corrections, which
have been calculated very recently \cite{KN1,Hbb}, will screen the leading
one--loop contribution by a non negligible amount. \s

This is, so far, for the SM Higgs particle. However, many extensions of the
Standard Model predict the existence of a larger Higgs sector. For instance
supersymmetric theories (SUSY), which are very attractive since at low energies
they provide a theoretical framework in which the problem of naturalness and
hierarchy in the Higgs sector is solved while retaining Higgs bosons with
moderate masses as elementary particles, require the existence of at least two
isodoublet scalar fields $\Phi_1$ and $\Phi_2$ to give masses separately to
isospin up and down particles, thus extending the physical spectrum of scalar
particles  to five \cite{HHG}. The physical Higgs bosons introduced by the
minimal supersymmetric extension of the Standard Model (MSSM) are of the
following type: two CP--even neutral bosons $h$ and $H$ [where $h$ will be the
lightest particle], a CP--odd neutral boson $A$ [usually called pseudoscalar]
and two charged Higgs bosons $H^{\pm}$. Besides the four masses $M_h$, $M_H$,
$M_A$ and $M_{H^\pm}$, two additional parameters define the properties of the
scalar particles and their interactions with gauge bosons and fermions: the
ratio of the two vacuum expectation values tg$\beta= v_2/v_1$ and a mixing
angle $\alpha$ in the neutral CP--even sector. Note that, contrary to a general
two--Higgs doublet model where the six parameters are free, supersymmetry leads
to several relations among these parameters and, in fact, only two of them are
independent\footnote{Note that there are also large radiative corrections to
the
supersymmetric Higgs boson masses and couplings \cite{SUSYRC}; these
corrections are beyond the scope of this paper and will not be discussed
here.}. \s

In this paper we calculate, by directly evaluating the relevant Feynam diagrams
using dimensional regularization, the fermionic contributions to the Higgs
boson self--energies at ${\cal O} (\alpha \alpha_S)$. To treat on the same
footing the case of scalar, pseudoscalar and charged Higgs particles, we have
considered the most general case where, in addition to leaving the momentum
tranfer arbitrary, we allow the internal quarks to be of different flavors $ U
\neq D$ and therefore to have different masses $m_U \neq m_D$. Our motivations
for performing such a calculation are threefold: \s

(i) As in the case of the SM Higgs boson, the strong \cite{RS,RA,RRY,RC} and
some of the electroweak \cite{EWSUSY} radiative corrections to SUSY Higgs boson
decays are known at the one--loop level. In some limiting cases, as for the
gluonic corrections for nearly massless quarks, SM two--loop results can be
adapted to the SUSY case \cite{RS3}. Here, we provide the necessary material
which allows to derive the universal part of the mixed ${\cal O}(\alpha_S G_F)$
radiative corrections to these Higgs decays. This is a generalization to a
multi--Higgs doublet model of the recent SM calculation \cite{KN1}, which is
just a special case [in the limit $m_U = m_D$ and when the Higgs boson is
CP--even] of the present results. \s

(ii) The imaginary parts of the Higgs boson self--energies are related, through
the optical theorem, to the hadronic\footnote{By hadronic Higgs decays, we mean
here the decays into quark pairs; the QCD corrections to the gluonic neutral
Higgs boson decays have been evaluated in Ref.~\cite{Hgg}.} partial decay
widths of the Higgs bosons. We give here the exact analytical expression of the
QCD corrections to the Higgs decay widths in the most general case $m_U \neq
m_D$ which is not available in the literature\footnote{QCD corrections to the
charged Higgs decay for $m_U \neq m_D$ have been calculated in Ref.~\cite{RC},
but not in a full analytical form since the integrals for the real corrections
have been performed numerically.}. In the special cases $m_U=m_D$, we recover
the known results for the QCD corrections to CP--even and CP--odd
\cite{RS,RA,RRY} neutral Higgs bosons which have been  obtained by directly
evaluating
the relevant Feynman amplitudes. Since our expressions have been obtained with
a completely different method, this serves as an independent check of these
results. \s

(iii) The fermionic contributions to the transverse and longitudinal components
of the electroweak vector boson self--energies at ${\cal O}(\alpha \alpha_S)$
have been evaluated recently in the general case \cite{nous}. It is known that
the longitudinal parts of the latter self--energies are directly related to the
corresponding Goldstone boson self--energies through a Ward Identity. Here, we
will show explicitly that this is indeed the case also for the Higgs bosons and
therefore provide a very powerful check of both calculations. \s

The paper is organized as follows. In the next section we will first set the
notation and summarize the one--loop results which will be relevant to our
discussion. In section 3 we will give a few details on the calculation of the
two--loop Higgs boson self--energies and discuss the renormalization procedure.
The complete results for the Higgs boson two--point functions
will be given in section 4. In section 5, we will derive the imaginary parts of
the two--loop self--energies which correspond to the QCD corrections to the
decays of the Higgs bosons into quark--antiquark pairs. Section 6 will
summarize our results. Finally, in the Appendix, we discuss the Ward Identity
which allows to relate our results to the ones derived for the longitudinal
components of the electroweak vector bosons.

\subsection*{2.~Notation and one--loop results}

\renewcommand{\theequation}{2.\arabic{equation}}
\setcounter{equation}{0}

In this section we will first set the notation and for the sake of
completeness, we rederive some of the one--loop results which will be relevant
to our next discussion. We will closely follow the notation of
Refs.~\cite{nous,moi}. \s

The contribution of a quark loop to the self--energy of a scalar Higgs boson
$\Phi$ will be denoted by $\Pi^\Phi(s=q^2)$ where $q$ is the four--momentum
transfer. To treat the cases of neutral CP--even, neutral CP--odd and charged
Higgs bosons on the same footing, it is convenient to work in the general
situation where the internal quarks in the loop are of different flavor, and
thus have different masses. This will correspond to the case of a charged
Higgs boson which couples to an up--type and down--type quark, with masses $m_U
\neq m_D \neq 0$; the self--energies of neutral scalar and pseudoscalar Higgs
bosons will simply be special cases of the previous one. \s

The coupling of charged Higgs bosons to fermions is a P--violating mixture
of scalar and pseudoscalar couplings

\vspace*{-5mm}

\be
g(H^+ U\bar{D}) = i \left( G_F/\sqrt{2} \right)^{1/2} \left[ ~h_U
(1-\gamma_5 ) ~+ ~h_D (1+ \gamma_5 ) ~ \right]
\en
where in a (type II) two--Higgs doublet model \cite{HHG} such as in the MSSM,
\be
h_U= m_U/{\rm tg}\beta \hspace*{0.4cm} , \hspace*{0.9cm} h_D=m_D {\rm tg} \beta
\en
It is often convenient to use the scalar and pseudoscalar components of this
coupling
\be
v = h_D+h_U \hspace*{0.4cm} , \hspace*{0.4cm} a= h_D-h_U
\en
The couplings of scalar, that we will denote by $S$, and pseudoscalar $A$ Higgs
bosons take the general form
\be
g(S Q \bar{Q}) = - i \left( G_F \sqrt{2} \right)^{1/2} ~h_{Q} \hspace*{0.4cm}
\hspace*{0.9cm} g(A Q\bar{Q}) = - \left( G_F \sqrt{2} \right)^{1/2} \gamma_5
{}~h_{Q}
\en
In the SM, the reduced couplings $h_{Q}$ are just the quark masses; in the
the MSSM these couplings for $\Phi=S,A$ are given in Table 1 for $U$ and $D$
quarks.

\begin{center}
\begin{tabular}{|c||c|c|} \hline
& & \\
$\hspace{0.3cm} \Phi \hspace{0.3cm}$ & $h_{U}/m_U $ & $h_{D}/m_D$ \\
& & \\ \hline \hline & & \\
$H_{SM}$ & 1  & 1  \\
$h$  & \ $\; \cos\alpha/\sin\beta \; $ \ & \ $ \; -\sin\alpha/\cos\beta \; $ \\
$H$  & \ $\; \sin\alpha/\sin\beta \; $ \ & $ \; \cos\alpha/\cos\beta $\ \\
$A$  & \ $\; 1/ {\rm tg}\beta  \; $& \ $ \; {\rm tg}\beta \; $ \\
& & \\ \hline
\end{tabular}

\vspace* {0.5cm}

{\small Tab.~1: Neutral Higgs couplings to up--type and down--type fermions
in the SM and MSSM.}
\end{center}

In the one--loop approximation, the contribution of a quark loop to the vacuum
polarization amplitude of a charged Higgs boson, $\Pi^C (q^2)$, will correspond
to $-i$ times the standard Feynman amplitude of the diagram Fig.~1a. For
arbitrary fermion masses $m_U \neq m_D \neq 0$ and momentum transfer $q^2$,
this amplitude reads
\be
\Pi^C (q^2)= \frac{N_CG_F}{\sqrt{2}} \left( \frac{\mu^2 e^\gamma}{4 \pi}\right)
^\epsilon \ i \int \frac{d^n k}{(2\pi)^n} {\rm Tr} \ \frac{(\nt k + m_U) (v -
a \gamma_5) (\nt k - \nt q +m_D) (v +a \gamma_5)}{(k^2-m_U^2)[(k-q)^2-m_D^2]}
\en
where $N_C=3$ is a color factor, $\mu$ is the 't Hooft renormalization mass
scale introduced to make the coupling constant dimensionless in $n=4-2\epsilon$
dimensions; we have also introduced an extra term $(e^\gamma /4\pi)^\epsilon$
[$\gamma$ is the Euler constant] to prevent uninteresting combinations of
$\log4\pi, \ \gamma \cdots$, in the final result. After calculating the trace
and integrating over the loop momentum, one obtains
\be
\Pi^C (s) = \frac{3 G_F}{2 \sqrt{2} \pi^2} \, s \, \left[~h_U^2 \Pi_U^+ (s) +
h_D^2 \Pi_D^+ (s) + 2 h_U h_D \frac{m_U m_D}{s} \Pi^- (s)~\right]
\en
where, in the general case $m_U \neq m_D \neq 0$, $\Pi^\pm (s)$ take the
following form
\be
\Pi^+_{U,D} (s) &=& \frac{1}{2\epsilon} (1+ 2\alpha +2 \beta) - \frac{1}{4}
(\rho_a +\rho_b) ( 1+ \alpha+\beta) - \frac{1}{2}\alpha \rho_a -\frac{1}{2}
\beta  \rho_b +1+ \frac{3}{2} \alpha+ \frac{3}{2} \beta \non \\
&+& \frac{1}{4} (1+ \alpha +\beta) \left[ (\alpha-\beta +\lam ) \log x_a
+(\beta-\alpha +\lam ) \log x_b \right] \non \\
\Pi^- (s) &=& - \frac{1}{\epsilon} -2 + \frac{1}{2} (\rho_a +\rho_b) -
\frac{1}{2} \left[ (\alpha-\beta +\lam ) \log x_a +(\beta-\alpha +\lam )
\log x_b \right]
\en
and where we use the variables [~$\beta = -m_D^2/s , \rho_b$ and $x_b$ are
defined in a similar way]
\be \alpha =-{m_{U}^2\over s} \ \ , \ \ \rho_{a}=\log{m_{U}^2\over \mu^2} \ \ ,
\ \ \ \ x_a =\frac{2 \alpha}{1+\alpha+\beta+\lam } \non
\en
\vspace*{-4mm}
\be
\lambda= 1+2\alpha +2\beta+ (\alpha -\beta)^2
\en

In the limit where one of the internal quarks is nearly massless compared to
its partner, as is the case for the top--bottom isodoublet, the coefficients
in front of $\Pi_D^+(s)$ and $\Pi^-(s)$ vanish and the expression of
$\Pi_U^+(s)$ simplifies to
\be
\Pi_U^+ (s) = \frac{1}{2} \left( \frac{1}{\epsilon}-\rho_a \right)
(1+2\alpha) +1+\frac{3}{2} \alpha +\frac{1}{2} (1+ \alpha)^2 \log \frac{\alpha}
{1+\alpha}
\en
The expressions of the self--energies for neutral scalar and pseudoscalar
Higgs bosons can simply be obtained by setting $m_U=m_D=m_Q$ in eqs.~(2.6--2.7)
and by using the relevant couplings which are given in Tab.~1. With the help of
the variable $x=4\alpha/(1+\sqrt{1+4\alpha})^2$ with $\alpha=-m_Q^2/s$, one has
\be
\Pi^{S,A} (s) = \frac{3 G_F}{2 \sqrt{2} \pi^2}\, s\, h_Q^2\, \left[~\Pi_Q^+(s)
\ \pm \ (m_Q^2/s)~\Pi^- (s) \right]
\en
leading to
\be
\Pi^S (s) &=& \frac{3 G_F}{2 \sqrt{2} \pi^2} \, s\, h_Q^2 \ \left[ \frac{1}{2}
\left( \frac{1}{\epsilon}-\rho_a \right) (1+6 \alpha) + 1+5\alpha +\frac{1}{2}
(1+4 \alpha)^{\frac{3}{2}} \log x \right] \non \\
\Pi^A (s) &=& \frac{3 G_F}{2 \sqrt{2} \pi^2} \, s\, h_Q^2 \left[ \frac{1}{2}
\left( \frac{1}{\epsilon}-\rho_a \right) (1+2 \alpha) +1+\alpha +\frac{1}{2}
(1+4 \alpha)^{\frac{1}{2}} \log x \right]
\en
Note that eq.~(2.6) exhibits the fact that $\Pi^{A,S}(s)$ can be obtained
from $\Pi^{S,A}(s)$ by simply making the substitution $m_U(m_D) \rightarrow
-m_U(-m_D$) in $\Pi^C (s)$ as expected from $\gamma_5$ reflection symmetry. \s

Finally, in the limit where the momentum squared is much larger or much smaller
than the quark masses squared, the self--energies read
\be
s \Pi^+(s \ll m_{U,D}^2) &=& - (m_U^2+m_D^2) \left( \frac{1}{\epsilon}
- \log \frac{m_U m_D}{\mu^2} +1 \right) +\frac{1}{2} \frac{m_U^4+m_D^4}
{m_U^2-m_D^2} \log \frac{m_U^2}{m_D^2} \non \\
\Pi^-(s \ll m_{U,D}^2 ) &=& - \frac{1}{\epsilon} + \log \frac{m_U m_D}{\mu^2}
-1 +\frac{1}{2} \frac{m_U^2+m_D^2}{m_U^2-m_D^2} \log \frac{m_U^2}{m_D^2}
\en
\be
s \Pi^+(s \gg m_{U,D}^2) = \frac{1}{2 \epsilon} - \frac{1}{2}
\log \frac{-s}{\mu^2} +1
\en

\vspace*{0.1cm}

In all the previous expressions the momentum transfer has been defined to
be in the space--like region, i.e. $s<0$. When continued to the physical region
above the threshold for the production of two fermions, $s \geq (m_U+m_D)^2$,
the Higgs boson self--energies acquire imaginary parts. These imaginary parts
are related to the decay widths of the Higgs particles into quark--antiquark
pairs. Adding a small imaginary part $-i \epsilon$ to the fermion masses
squared, the analytical continuation is consistently defined.

\ From the expressions eqs.~(2.7), the imaginary parts can be straightforwardly
obtained by making the substitution
\be
\log x_{a,b} \rightarrow \log|x_{a,b}| + i \pi
\en
One then has for the partial decay widths of a charged Higgs boson $H^+$ into
$U \bar{D}$ quark pairs [$s=M_{H^+}^2$]
\be
\Gamma (H^+\rightarrow U\bar{D})=\frac{N_CG_FM_{H^+}}{2 \sqrt{2} \pi^2}
\left[h_U^2 {\cal I}m \Pi_U^+ (s) + h_D^2 {\cal I}m \Pi_D^+ (s) + 2 h_U h_D
\frac{m_U m_D}{s} {\cal I}m \Pi^- (s)\right]
\en
with
\vspace*{-5mm}
\be
{\cal I} m\Pi^+_{U,D} (s) &=& \frac{\pi}{2} \lam \left( 1+ \alpha +\beta
\right) \hspace*{0.5cm} , \hspace*{0.7cm} {\cal I}m\Pi^- (s) = -\pi \lam
\en
In the limit $m_D=0$ this partial decay width reduces to the more familiar form
\be
\Gamma (H^+ \rightarrow U\bar{D}) = \frac{N_C G_F}{4 \sqrt{2} \pi} M_{H^+}
\ h_U^2 ~ \left( 1- \frac{m_U^2}{s} \right)^2
\en

In a similar manner, one also obtains the familiar expressions of the
partial decay widths of neutral scalar and pseudoscalar Higgs bosons into
quark--antiquark pairs
\be
\Gamma (S \rightarrow Q \bar{Q} ) &=& \frac{N_C G_F}{4 \sqrt{2} \pi} \, M_S \,
h_Q^2 \ \left( 1 - 4 \frac{m_Q^2}{s} \right)^{3/2} \non \\
\Gamma (A \rightarrow Q \bar{Q} ) &=& \frac{N_C G_F}{4 \sqrt{2} \pi} \, M_A \,
h_Q^2 \ \left( 1 - 4 \frac{m_Q^2}{s} \right)^{1/2}
\en

\subsection*{3.~Two--loop calculation}

\renewcommand{\theequation}{3.\arabic{equation}}
\setcounter{equation}{0}

At ${\cal O}(\alpha \alpha_S)$, the two--loop diagrams contributing to the
Higgs boson self--energies $\Pi^\Phi(q^2)$ [up to a factor $-i$] are shown in
Fig.~1b. In the 't Hooft--Feynman gauge, using the routing of momenta shown in
the figure and following the notations introduced in the previous section, one
can write the bare amplitude as
\be
\left. \Pi^\Phi(q^2) \right|_{\rm bare}=  \frac{16 \pi G_F }{3 \sqrt{2}} N_C
\alpha_S \, \left( \frac{\mu^2 e^\gamma}{4\pi} \right)^{2\epsilon} \, \int
\frac{d^nk_1}{(2\pi)^n} \ \int \frac{d^n k_2}{(2\pi)^n} \ \left[ {\cal A}
+{\cal B} \right]
\en
where, for $m_U \neq m_D \neq 0$, are given by
\be
{\cal A} &=& {\rm Tr} \frac{ (\nt k_1+m_U) (v - a\gamma_5) (\nt k_1 -\nt q
+m_D) \gamma_\lambda (\nt k_2 -\nt q +m_D) (v+ a \gamma_5) (\nt k_2+m_U)
\gamma^\lambda }{ (k_1-k_2)^2 \ (k_1^2-m_U^2)
\ (k_2^2-m_U^2) \ [(k_1-q)^2-m_D^2] \ [(k_2-q)^2-m_D^2]} \non \\
{\cal B} &=& {\rm Tr} \frac{ (\nt k_1+m_U)(v-a\gamma_5) (\nt k_1 -\nt q +m_D)
(v+a \gamma_5) (\nt k_1+m_U) \gamma_\lambda (\nt k_2 +m_U) \gamma^\lambda}{
(k_1-k_2)^2 \ (k_1^2-m_U^2)^2 \ (k_2^2-m_U^2) \ [(k_1-q)^2-m_D^2] } \non \\
& & \ + \ m_U \ \longleftrightarrow m_D \
\en

This bare amplitude has to be supplemented by counterterms; these include the
quark wave function and mass counterterms as well as the Higgs--quarks vertex
counterterm. However, the renormalization of the Higgs--quark vertex is
connected with the renormalization of the quark masses and wave
functions; because the latter counterterms cancel, one only needs to include
quark mass renormalization and considers the diagrams where the quark mass
counterterms are inserted into the one--loop Higgs boson self--energy; Fig.~2a.
The mass counterterm is obtained by evaluating the amplitude of the diagram
shown in Fig.~2b, which reads in dimensional regularization
\be
-i \Sigma(\nt p) = - \alpha_s \frac{16 \pi}{3} \left( \frac{e^\gamma \mu^2 }
{4\pi} \right)^\epsilon \int \frac{d^nk}{(2\pi)^n} \frac{ \gamma^\lambda (\nt p
- \nt k + m_Q) \gamma_\lambda } {[(p-k)^2-m^2_Q] \ k^2}
\en
where $p$ is the four--momentum of the quark and $m_Q$ its bare mass. This
expression can be decomposed into a piece proportional to $(\nt p-m_Q)$ which
will enter the wave function renormalization and another piece proportional
to $m_Q$ which will give the mass counterterm. After integration over the loop
momentum, the latter is given by
\be
\Sigma (p^2) = \frac{\alpha_s}{\pi} \frac{m_Q}{\epsilon} \left(
\frac{e^\gamma \mu^2}{m^2_Q} \right)^{\epsilon} \frac{1-2/3 \epsilon}{1-
2\epsilon} \Gamma (1+\epsilon) \ \left[1 +{\cal O}(p^2/m_Q^2-1) \right]
\en
The mass counterterm will now depend on the renormalization procedure, i.e. on
the way the quark mass is defined. In the on--shell scheme which is usually
used to calculate radiative corrections in the electroweak theory \cite{sir},
the fermion masses are defined at $p^2=m^2_Q$ and correspond to the position of
the pole of the fermion propagators; they are referred to as the on--shell
masses and the counterterm reads in this case\footnote{ Note that one can also
employ a different definition of the quark masses; for instance one can use the
$\overline{\rm MS}$ mass which is defined by just picking the divergent term in
the expression of $\Sigma(p^2)$ in eq.~(3.4), or the running mass where one
evaluates $\Sigma(p^2)$ at a scale $p^2=M_\Phi^2$. Having at hand the
expressions of
the two--loop self--energies in the on--shell scheme, the procedure for
obtaining the corresponding result in any other renormalization scheme is
straightforward and can be found, e.g., in Ref.~\cite{nous}. In practice,
however, it is sufficient to replace in the one--loop result, the on--shell
mass by the $\overline{\rm MS}$ or the running mass: the difference between
this
result and the one obtained using the procedure discussed in Ref.~\cite{nous}
is of higher order in $\alpha_S$.}
\be
\delta m_Q \ \equiv  \ m_Q (m^2_Q) - m_Q \ = \ \frac{\alpha_S}{\pi} \frac{m_Q}
{\epsilon} \left( \frac{\mu^2}{m^2_Q}\right)^\epsilon \ \left( 1+ \frac{\pi^2}
{12}\epsilon^2 \right) \ \frac{1-2\epsilon/3}{1-2\epsilon}
\en
One then inserts this mass counterterm in the one--loop self--energies, as
depicted in the diagrams of Fig.~2a, which is equivalent to calculate
\be
\left. \Pi^\Phi (q^2) \right|_{\rm CT} = \ - \delta m_U \left[ 1+ \frac{
\partial}{\partial m_U} \right] \left. \Pi^\Phi (q^2) \right|_{\rm 1-loop}
\ - \delta m_D \left[ 1+ \frac{\partial}{\partial m_D } \right] \left. \Pi^\Phi
(q^2)\right|_{\rm 1-loop}
\en
where the one--loop vacuum polarization function is given by eq.~(2.6--2.7),
up to ${\cal O}(\epsilon)$ terms which have to be included. The renormalized
two--loop self--energies will then read
\be
\Pi^\Phi (q^2)=\left. \Pi^\Phi(q^2)\right|_{\rm bare}+\left. \Pi^\Phi(q^2)
\right|_{\rm
CT}
\en

Similarly to the one--loop case, after calculating the trace in eqs.~(3.2)
and expressing the scalar products of the momenta appearing in the numerators
in terms of combinations of the polynomials in the denominators, one is led to
the calculation of a set of scalar two--loop integrals \cite{bro} which can
also be found in Ref.~\cite{nous}. In the following we will not give any more
details on the cumbersome calculation: we will simply list our main
results in the next two sections.

\subsection*{4.~Two--loop self--energies}
\renewcommand{\theequation}{4.\arabic{equation}}
\setcounter{equation}{0}

We begin by giving the expression of the contribution of a $(U,D)$ isodoublet
to the charged Higgs boson two--point function at order ${\cal O}(\alpha
\alpha_S)$ in the general case $m_U \neq m_D \neq 0$. The result will be given
in the on--shell mass scheme; i.e. $m_{U,D}$ will stand for the on--shell
masses. Using the same notations as in the one loop case [confusion should be
rare], the charged Higgs boson self--energy $\Pi^C(q^2)$ at the two--loop level
is given by
\be
\Pi^C (s)= \frac{G_F}{2 \sqrt{2} \pi^2} \frac{\alpha_S}{\pi} \ s \ \left[ \
h_U^2 \Pi^+_U (s) + h_D^2 \Pi^+_D (s) \ + \ 2 h_U h_D \ \frac{m_U m_D}{s} \
\Pi^- (s)\ \right]
\en
with $\Pi^\pm (s)$ given by the relatively simple and compact expressions
\be
\Pi^+_U &=& - \frac{3}{2\epsilon^2}(1+4\alpha+4 \beta) - \frac{1}
{\epsilon} \left[ \frac{11}{4}+ 14 \alpha + 14 \beta -3 \rho_a -12 \alpha\rho_a
-6 \beta \rho_a -6 \beta \rho_b \right] \non \\
&+& (\rho_a +\rho_b) \left[\frac{11}{4}+14 \alpha +14\beta  - 3(\rho_a+\rho_b)
\left( \frac{1}{4} +\alpha+ \beta \right) - 3 (\alpha- \beta) (\rho_a -\rho_b)
\right] \non \\
&+& (\rho_a -\rho_b) \left[5 + \frac{17}{2} \alpha +\frac{17}{2} \beta
-\frac{3}{2} (1+2 \alpha+ 2 \beta) (\rho_a+ \rho_b) + \frac{3}{4} (\alpha^2
-\beta^2)(\rho_a -\rho_b) \right] \non \\
&+& \frac{3}{8} - \frac{53}{2} (\alpha +\beta) -\frac{\pi^2}{4}(1+4\alpha
+4\beta) + \frac{3}{4} \lam (1+ \alpha +\beta) (\log x_a + \log x_b )
(\rho_a-\rho_b) \non \\
&+ & \frac{1}{4} \log x_a \left[ 31(\alpha -\beta) +9(1+\alpha +\beta)
(\alpha-\beta+\lam)\right]
+ \frac{1}{4} \log ^2 x_a \left[ (1-2\alpha-2\beta) \right. \non \\
& \times & \left. \left( -\lambda+1
+\alpha + \beta -(\alpha-\beta) \lam \right) -3 (\alpha+\beta) +3(\alpha -
\beta ) (\alpha -\beta +\lam) \right] \non \\
&+& \frac{1}{4} \log x_b \left[ 31(\beta-\alpha ) +9(1+\alpha +\beta)
(\beta- \alpha +\lam) \right]
+ \frac{1}{4} \log^2 x_b \left[ (1-2\alpha-2\beta) \right. \non \\
& \times & \left. \left( -\lambda+1
+\alpha + \beta +(\alpha-\beta) \lam \right) -3 (\alpha+\beta) +3(\alpha -
\beta ) (\alpha -\beta - \lam) \right] \non \\
&+& \frac{3}{2} \log x_a \log x_b (\lambda +2 \alpha + 2\beta + 6 \alpha \beta)
-(1+ \alpha+ \beta)^2 {\cal I} -2(1+\alpha+\beta) {\cal I}'
\en
\be
\Pi^+_D \ \ = \ \ \Pi_U^+ \ [ \ m_U \longleftrightarrow \ m_D \ ]
\en
\be
\Pi^- &=& \frac{6}{\epsilon^2} +\frac{1}{\epsilon} ( 14 -6\rho_a -6\rho_b)
- 14 \rho_a  - 14 \rho_b + 3 (\rho_a +\rho_b)^2 +20 + \pi^2 \non \\
&-& 6 \log x_a (\alpha-\beta + \lam ) - \log ^2 x_a \left[ \lambda -1 -\alpha
-\beta +(\alpha- \beta) \lam \right] \non \\
&- & 6 \log x_b (\beta -\alpha + \lam ) - \log ^2 x_b \left[ \lambda -1 -\alpha
-\beta +(\beta - \alpha ) \lam \right] \non \\
& - & 6 \log x_a \log x_b (1+\alpha+\beta) +2(1+\alpha+\beta) {\cal I} + 4{\cal
I}'
\en

\vspace*{0.1cm}

In these expressions, ${\cal I}$ and ${\cal I}'$ given by
\be
{\cal I} & = & F(1)+F(x_a x_b)-F(x_a)-F(x_b) \non \\
{\cal I}' &= & \lam G(x_a x_b) -\frac{1}{2} (\beta-\alpha+\lam )G(x_a)
-\frac{1}{2}( \alpha - \beta + \lam ) G(x_b)
\en
where in terms of the polylogarithmic functions \cite{levin} ${\rm
Li}_2(x)= -\int_0^1 y^{-1}\log (1-xy) {\rm d}y$ and ${\rm Li}_3(x)= -\int_0^1
y^{-1} \log y \log(1-xy)  {\rm d}y$, the functions $F$ and $G$ are given by
\be
F(x) & = & 6{\rm Li}_3(x)-4{\rm Li}_2(x) \log x - \log^2 x \log (1-x) \non \\
G(x) &= & 2{\rm Li}_2(x) +2 \log x \log (1-x)+ \frac{x}{1-x} \log^2 x
\en

\vspace*{0.1cm}

In the limit where one of the quarks is massless, $m_D=0$, the
coefficients of $\Pi^+_D$ and $\Pi^-$ vanish while $\Pi_U^+$ takes the
much simpler form [$x= \alpha/(1+\alpha)$ with $\alpha=-m_U^2/s$]
\be
\Pi^+_U (s)&=& - \frac{3}{2\epsilon^2} (1+4\alpha) - \frac{1}{\epsilon} \left[
{11 \over 4} + 14 \alpha - 3 \rho_a - 12 \alpha \rho_a \right] +\frac{1}{2}
\rho_a (11+ 56 \alpha ) \non \\
&- & 3\rho_a^2 (1+4 \alpha) +\frac{3}{8} -\frac{53}{2} \alpha- \frac{\pi^2}{4}
(1+4\alpha) +\frac{9}{2} (1+\alpha)^2 \log x \non \\
&+&  \frac{1}{2} (1+\alpha)^2 (3+2 \alpha) \log^2 x +(1+\alpha)^2 [F(x)
- F(1)] + (1+\alpha) G(x)
\en

Note that in this limit, the expression of $\Pi_U^+ (s)$ is free of mass
singularities as it should be. \s

As in the one--loop case, one can derive the expressions of the self--energies
for neutral scalar and pseudoscalar Higgs bosons from eqs.~(4.1-4.4) by
setting $m_U=m_D=m_Q$ and using the proper couplings. One would have [$x =4
\alpha / (1+\sqrt{1+4\alpha} )^2$ with $\alpha=-m_Q^2/s$]
\be
\Pi^{S(A)}_Q (s) = \frac{G_F}{2 \sqrt{2} \pi^2} \ \frac{\alpha_S}{\pi} \ s \
h_Q^2 \ S_Q (A_Q)
\en

\nn with $S_Q=\Pi_Q^+(s)-\alpha\Pi^-(s)$ and $A_Q=\Pi_Q^+(s)+\alpha\Pi^-(s)$
given by
\be
S_Q \hspace*{-0.2cm} &=& \hspace*{-0.2cm} - \frac{3}{2 \epsilon^2} (1+12
\alpha) - \frac{1}{
\epsilon} \left(\frac{11}{4} - 3 \rho_a +42 \alpha- 36 \alpha \rho_a \right)
+ \frac{11}{2} \rho_a - 3 \rho_a^2 +84 \alpha \rho_a -36 \alpha \rho_a^2 \non
\\
&+ & \frac{3}{8} -73 \alpha -\frac{\pi^2}{4}( 1+12\alpha) + \frac{3}{2}
(1+4\alpha)^{\frac{1}{2}} (14\alpha +3) \log x + (\frac{3}{2}+14\alpha+
29\alpha^2) \log ^2 x \non \\
&-& (1+ 2\alpha) (1+ 4\alpha) \left[ F(1)+F(x^2)-2F(x) \right] - 2
(1+4\alpha)^{\frac{3}{2}} \left[ G(x^2)-G(x) \right] \non \\
A_Q \hspace*{-0.2cm} & = & \hspace*{-0.2cm} -\frac{3}{2 \epsilon^2} (1+4
\alpha) - \frac{1}{
\epsilon} \left( \frac{11}{4} - 3 \rho_a +14 \alpha- 12 \alpha \rho_a \right)
+\frac{11}{2} \rho_a - 3 \rho_a^2 + 28 \alpha \rho_a -12 \alpha \rho_a^2 \non
\\
&+& \frac{3}{8} -33 \alpha - \frac{\pi^2}{4}( 1+4\alpha) + \frac{3}{2} (1+4
\alpha)^{\frac{1}{2}}(3- 2\alpha)\log x +(\frac{3}{2}+2\alpha-3\alpha^2)\log
^2x \non \\
&-& (1+ 2\alpha) \left[ F(1)+F(x^2)-2F(x) \right] - 2 (1+4\alpha)^{\frac{1}{2}}
\left[ G(x^2)-G(x) \right]
\en
The expression of $\Pi^S_Q$ has been derived very recently \cite{KN1} in the
case of the Standard Model Higgs boson; we have verified that both results are
in agreement with each other\footnote{We have also found that the leading
${\cal O}(G_F \alpha_S m_t^2)$ universal radiative correction factor to the
Higgs boson fermionic decay widths, for $m_t \gg m_b$, is indeed $\left[1-
\frac{6}{7} (1+\pi^2/9) \frac{\alpha_S}{\pi} \right]$, in accord with
Ref.~\cite{KN1}. In the case of the pseudoscalar Higgs boson with tg$\beta=1$,
the correction factor in the limit $m_t \gg m_b$ is found to be $\left[1-
\frac{2}{9} (3+\pi^2) \frac{\alpha_S}{\pi} \right]$; this is just the
well--known QCD correction to the $\rho$ parameter \cite{moi}. This result is
not surprising if one recalls that the pseudoscalar Higgs boson couples like
the
longitudinal component of the $Z$ boson [see Appendix] which, in turn, is the
same as the transverse component for $q^2=0$. Note, however, that this
contribution is the full ${\cal O}(G_F\alpha_S m_t^2)$ radiative correction
only for the leptonic Higgs boson decays: for decays into quark pairs [and not
only in the case of the $b$ quark] additional diagrams have to be considered;
see also Ref.~\cite{Hbb}.}. \s

Finally, in the limit where the momentum squared is much smaller than the
internal quark masses squared, the components $\Pi^+_Q$ and $\Pi^-$ read
\be
s\Pi^+_Q &=& \frac{6}{\epsilon^2} m_+ + \frac{1}{\epsilon} \left(14 m_+ -3m_+
\rho_- -3 m_-\rho_- - 6m_+ \rho_+ \right) -\frac{3}{4} \frac{m_+^3}{m_-^2}
\rho_-^2 + \frac{3}{4} \frac{m_+^2}{m_-} \rho_-^2 +\pi^2 m_+ \non \\
&+ & m_+ \left( 3\rho_+^2 -14 \rho_+ -7 \rho_- +3 \rho_+ \rho_- +\frac{9}{4}
\rho_-^2 +30 \right)+ m_- \left( \frac{3}{4}\rho_-^2 -7 \rho_- + 3
\rho_+ \rho_- \right) \non \\
\Pi^- &=& \frac{6}{\epsilon^2} + \frac{1}{\epsilon} (14 -6\rho_+)
- \frac{3}{2} \frac{m_+^2}{m_-^2} \rho_-^2 + 30 -14 \rho_+ +3 \rho_+^2
+\pi^2 +\frac{3}{2}\rho_-^2
\en
with $\rho_\pm = \log\frac{m_U^2}{\mu^2} \pm \log\frac{m_D^2}{\mu^2}$ and
$m_\pm = m_U^2 \pm m_D^2$. In the opposite limit, i.e. when the masses are very
small compared to the momentum transfer squared, the coefficient of $\Pi^-$
vanishes and $\Pi^+_Q$ will read [$\zeta(3)=F(1)/6=1.202$]
\be
s\Pi^+_Q &=& -\frac{3}{2\epsilon^2}+\frac{1}{\epsilon} \left(3
\log\frac{m_Q^2}{\mu^2} -\frac{11}{4} \right) + \frac{3}{8}-\frac{\pi^2}{4}
- 6\zeta(3) \non \\
&+& 10 \log\frac{m_Q^2}{\mu^2} -3 \log^2\frac{m_Q^2}{\mu^2}
-\frac{9}{2} \log\frac{-s}{\mu^2}  + \frac{3}{2} \log^2 \frac{m_Q^2}{-s}
\en

\subsection*{5.~Hadronic decay widths}
\renewcommand{\theequation}{5.\arabic{equation}}
\setcounter{equation}{0}

We now turn to the discussion of the partial decay widths of these various
Higgs
bosons in quark--antiquark pairs. At ${\cal O}(\alpha_S)$, the partial decay
width of a charged Higgs boson into $U\bar{D}$ pairs is given by
$[s=M_{H^+}^2]$
\be
\Gamma (H^+ \rightarrow U \bar{D}) = \frac{G_F\alpha_S M_{H^+}}{2 \sqrt{2}
\pi^3} \left[h_U^2 {\cal I}m \Pi_U^+ (s) + h_D^2 {\cal I}m \Pi_D^+ (s)
+ 2 h_U h_D \frac{m_U m_D}{s} {\cal I}m \Pi^- (s) \right]
\en

The imaginary parts of $\Pi^+_{U,D}$ and $\Pi^-$ can be derived along the same
lines as discussed previously in the one--loop case. Using the fact
that
\be
{\cal I}m \log x_{a,b} = \pi \hspace*{0.5cm} , \hspace*{1cm}
{\cal I}m \log^2 x_{a,b}  = 2\pi \log|x_{a,b}| \non
\en
\vspace*{-5mm}
\be
{\cal I}m{\cal I}' \hspace*{-.2cm} &= & \hspace*{-.2cm} \pi {\cal J}' = \pi
\left\{ 4 \lam \left[ \log (1-x_ax_b)+\frac{x_ax_b}{1-x_ax_b} \log|x_a x_b|
\right] -(\beta-\alpha+ \lam ) \left[ \log (1-x_a) \right. \right. \non \\
& + & \left. \left. \frac{x_a}{1-x_a} \log |x_a|\right]
- (\alpha-\beta+ \lam ) \left[ \log (1-x_b)+\frac{x_b}{1-x_b} \log |x_b|
\right] \right\} \non
\en
\be
{\cal I}m{\cal I}  &= & \pi {\cal J} \ = \ -2 \pi \left[ 4 {\rm
Li}_2 (x_ax_b) -2{\rm Li}_2(x_a)-2{\rm Li}_2(x_b) +2\log|x_a x_b|
\right. \non \\
& \times & \left. \log (1-x_ax_b) -\log|x_a| \log (1-x_a) -
\log|x_b| \log(1-x_b) \right ]
\en

\nn one obtains for ${\cal I}m \Pi^\pm$ in the general case $m_U \neq m_D\neq
0$
\be
\frac{1}{\pi} {\cal I}m \Pi^+_U(s) &=&
\left[ (1+\alpha+\beta) \left( \alpha-\beta + \frac{3}{2} \right) \lam
+ \left( \frac{3}{2} +\alpha+\beta \right) \lambda +5 \alpha \beta \right]
\log |x_a| \non \\
&+ & \left[ (1+\alpha+\beta) \left( \beta - \alpha -\frac{3}{2} \right) \lam +
\left( \frac{3}{2} +\alpha+\beta \right) \lambda +5 \alpha \beta \right]
\log |x_b| \non \\
&+& \frac{9}{2}(1+\alpha+\beta)\lam -  (1+\alpha+\beta)^2 {\cal J} -
2 (1+\alpha+\beta) {\cal J}'
\en
\be
{\cal I}m \Pi^+_D(s) \ = \ {\cal I}m \Pi^+_U(s) \ [ \ m_U \longleftrightarrow
m_D \ ]
\en
\be
\frac{1}{\pi} {\cal I}m \Pi^-(s)
&= & - 2 \left[ \lambda+ 2(1+\alpha+\beta) + (\alpha -\beta ) \lam \right]
\log |x_a| \non \\
& & - 2 \left[ \lambda+ 2(1+\alpha+\beta) + (\beta- \alpha) \lam \right]
\log |x_b| \non \\
& & -12 \lam + 2(1+\alpha+\beta) {\cal J} + 4 {\cal J}'
\en

\ From this formulae one can derive again the expressions of the hadronic decay
widths in the previous special situations of physical relevance. In the limit
where one of the quark is nearly massless, $m_D \rightarrow 0$, one has for
${\cal I}m \Pi^+_U(s)$
\be
C_Q= \frac{1}{\pi } {\cal I}m \Pi^+_U (s) & = & \frac{9}{2} (1+\alpha)^2
+ (1+\alpha)(3+7\alpha+2 \alpha^2) \log \frac{-\alpha}{1+\alpha}
- 2 (1+\alpha)^{2} \non \\
& \times &  \left[ \frac{\log (1+\alpha)}{1+\alpha} + 2{\rm Li
}_2 \left( \alpha\over \alpha +1 \right)- \log (1+\alpha) \log{-\alpha \over
1+\alpha} \right]
\en

In the case of scalar and pseudoscalar Higgs bosons, the partial decay widths
$\Gamma (S,A \rightarrow Q \bar{Q})$ will be given by
\be
\Gamma[S(A) \ra Q \bar{Q}]= \frac{ G_F}{2 \sqrt{2} \pi^2} \frac{\alpha_S} {\pi}
h_Q^2 M_{S(A)} \ S_Q (A_Q)
\en
where $S_Q / A_Q= {\cal I}m \Pi_Q^+(s, \beta=\alpha)/\pi \mp \alpha {\cal I}m
\Pi^-(s,\beta=\alpha)/\pi$ are given by
\be
S_Q &=& \frac{3}{2} (1+4 \alpha)^{\frac{1}{2}} (14\alpha+3) + (58\alpha^2
+28\alpha+3) \log |x| - 4 (1+4 \alpha) J \non \\
A_Q &=& \frac{3}{2} (1+4 \alpha)^{\frac{1}{2}} (3-2\alpha ) + (3+4\alpha
-6\alpha^2) \log |x| - 4J
\en
where

\vspace*{-5mm}

\be
J &= & - (1+2\alpha) \left[ 2{\rm Li}_2 (x^2) - 2{\rm Li}_2(x)
+2\log|x| \log (1-x^2) - \log|x| \log (1-x) \right] \non \\
&& +  2\sqrt{1+4\alpha} \left[ 2 \log (1-x^2) - \log (1-x)
+\frac{x(3x -1)}{1-x^2} \log|x| \right]
\en

Finally, let us note that in the limit where the quark masses are much
smaller than the Higgs boson masses, the QCD corrections to the decay widths
will exhibit the well known logarithmic behavior \cite{RS,RA} which, because of
chiral symmetry, is the same for the scalar, pseudoscalar and charged Higgs
boson
\be
S_Q, A_Q, C_Q \ra \frac{9}{2} + 3 \log \frac{m^2_Q}{M_\Phi^2}
\en

One has therefore to sum these potentially large logartithmic terms; this is
equivalent to replace the on--shell quark masses by the running masses defined
at $p^2=M_\Phi^2$ when renormalizing the $\Phi Q \bar{Q}$ vertex. \s

Analytical results for ${\cal I}m \Pi^S(s)$ \cite{RS,RA,RRY} and ${\cal I}m
\Pi^A(s)$ \cite{RA,RRY} have been obtained in the past by a number of authors
by
directly calculating the QCD corrections to the decay of a scalar Higgs boson
into quark pairs. The results that we obtain here using a completely different
method agree with the previous ones; this serves as check of our full
calculation in the general case. Note also that for the value tg$\beta=1$,
we recover the expression of the imaginary part for the longitudinal
component of the electroweak vector bosons in the general case, which is
given in Refs.~\cite{RRY,nous,STN}. Indeed, because of a Ward identity
which will be discussed in the forthcoming Appendix, the imaginary part of the
longitudinal component of the vector boson self--energy is the same as the
one for the Higgs boson self--energy for this value of tg$\beta$. This feature
provides also a very poweful check of the calculation presented here.

\subsection*{6. Summary}

In this paper, the contribution of heavy quarks to the Higgs boson
self--energies were calculated at first order in the strong interaction. We
have considered the most general case: finite momentum transfer and arbitrary
masses for the internal quarks to treat on the same footing the case of scalar,
pseudoscalar and charged Higgs bosons; these particles appear in many
extensions of the Standard Model scalar sector such as two--Higgs doublet
models and in particular, supersymmetric theories. Full analytical formulae for
the real parts of the self--energies at ${\cal O} (\alpha_S)$ were presented in
the on--shell quark mass scheme. \s

We have also given the expressions of the self--energies in some situations of
physical interest: the case where the two quarks have equal masses which
corresponds to neutral scalar and pseudoscalar Higgs bosons, the case where
one of the quarks has a negligible mass with respect to the other which would
correspond to the approximate contribution of the top--bottom isodoublet to the
charged Higgs boson self--energy, and finally the case where the momentum
transfer squared is much larger or much smaller than the quark masses squared.
\s

By analytical continuation, we have then derived the imaginary part of the
Higgs boson self--energies in the general case $m_U \neq m_D \neq 0$; this
imaginary part corresponds to the QCD correction to the partial decay width of
the Higgs bosons into quark--antiquark pairs. In the special case $m_U =m_D$
these corrections have been obtained by several authors in a full analytical
form. In these limits, the results that we obtain here using a completely
different method provide independent checks of these calculations.  \s

Finally, in the forthcoming Appendix, we relate the results for the two--loop
Higgs boson self-energies that we obtained here by directly evaluating the
relevant Feynman amplitudes, to the results for the longitudinal components of
the electroweak vector boson vacuum polarization functions which are available
in the literature. This provides a consistency check of both calculations.

\vspace*{0.5cm}

\nn {\bf Acknowledgements.} \s

\nn Discussions with Alberto Sirlin and technical help from J. Papavassiliou
are gratefully acknowledged. One of us (PG) would like to thank the Groupe de
Physique des Particules of the Universit\'e de Montr\'eal for
the hospitality extended to him in the final stage of this work. This work is
partially supported by the National Sciences and Engineering Research Council
of Canada and by the National Science Foundation under Grant No. PHY--9313781.

\newpage

\begin{center} \section*{\bf APPENDIX} \end{center}

\begin{center}
\subsection*{Ward Identities and Goldstone boson self--energies.}
\end{center}

\renewcommand{\theequation}{A.\arabic{equation}}
\setcounter{equation}{0}

As is well--known, there are Ward identities relating the longitudinal
components of the electroweak vector bosons and the corresponding Goldstone
bosons \cite{WI}. In this Appendix, we use the current algebra of the Standard
Model to derive these Ward identities; we will then briefly show how to relate
the Higgs boson self--energies calculated in the previous sections to the
longitudinal parts of the electroweak vector boson self--energies, the
expressions of which have been derived up to ${\cal O}(\alpha \alpha_S)$ in the
general case in Ref.~\cite{nous}. \s

Defining the fermionic contribution to the vacuum polarization function of the
$W$ boson and of the corresponding Goldstone boson $\Phi$ as [$g$ is the
SU(2)$_L$ coupling constant]
\be
\Pi_{WW}^{\mu\nu}(q^2)= -i \frac{g^2}{2}\int d^n x e^{-i q\cdot x}
\langle 0|T^* J_W^{\dagger\mu}(x) J_W^\nu(0)|0\rangle
\en
\be
\Pi_{\Phi\Phi}(q^2)= + i \frac{g^2}{2 M_W^2}\int d^n x e^{-i q\cdot x}
\langle 0|T^* S^\dagger (x) S(0)|0\rangle
\en
where $J^\mu_W(x)$ and $S(x)$ are the charged fermionic currents coupled to the
$W$ and to the $\Phi$ bosons and T$^*$ denotes the covariant time ordering
product; for the notation and normalization of the currents, we will follow
Ref.~\cite{SIR78}. Contracting $\Pi^{\mu \nu}_{WW}$ with the tensor $q^\mu
q^\nu$,
one obtains
\be
q^\mu q^\nu \int d^n x e^{-i q\cdot x}
\langle 0|T^* J_W^{\dagger \mu} (x) J_W^\nu(0)|0\rangle=\int d^n x e^{-i
q\cdot x} \langle 0|T^* S^\dagger(x) S(0)|0\rangle- \frac{i}{2}\langle
0|S_1(0)|0\rangle \hspace*{0.1cm}
\en
with $S_1$ the current coupled to the Standard Model Higgs boson and where
we have used
\be
\partial_\mu J_W^\mu (x) = i S(x)
\en
and
\be
[J^0_W(x), S^\dagger (y)]_{x^0=y^0}= +\frac{1}{2}\ \delta^3(\vec x-\vec y)
[S_1(x) -i S_2(x)]
\en
where $S_2$ is the current coupled to the neutral Goldstone boson. This,
in turn, can be written as
\be
q^2 \Pi_{WW}^L(q^2)= - M_W^2 \Pi_{\Phi \Phi}(q^2) - \frac{g^2}{4}\langle
0| S_1(0)|0\rangle \label{wiz}
\en

This equation relates the longitudinal part of the vacuum polarization function
of the $W$ boson to the self--energy of the corresponding unphysical charged
boson. One can see that the subtraction term $\langle 0|S_1(0)|0 \rangle$
[a tadpole] is needed  to cancel a spurious quartic dependence on the mass
of the fermions. \s

Even though the previous derivation was at the one--loop level in the
electroweak interactions, it is valid at any order in the strong interactions
as the QCD generators commute with the ones of the electroweak group.
To derive the self--energy of the charged Goldstone boson at ${\cal O}(\alpha)$
and ${\cal O}(\alpha \alpha_S)$, we therefore need only the expressions of the
electroweak vector boson self--energies given in Ref.~\cite{nous} in the
general case and the one of the tadpole diagrams of Fig.~3 where both the
two quarks of the same weak isodoublet are running in the loop. Using the same
notation as in the main text, and  for a single quark of mass $m_Q$ [which is
renormalized ``on-shell"], we obtain for the tadpole amplitude up to order
$\alpha_S$
\be
\langle 0|S_1(0)|0\rangle &=&  \frac{3m^4_Q}{4\pi^2} \left[ \frac{1}{\epsilon}
+ 1 -  \log\frac{m_Q^2}{\mu^2} + \frac{\alpha_s}{3\pi} \left( -\frac{6}
{\epsilon^2} -\frac{1} {\epsilon} \left( 14-12 \log\frac{m_Q^2}{\mu^2} \right)
\right. \right. \non \\
& & \hspace*{2cm} \left.  \left. -30- \pi^2 +28 \log\frac{m_Q^2}{\mu^2}  -
12 \log^2\frac{m_Q^2}{\mu^2} \right) \right]
\label{tadpole}
\en

This equation, added to the one and two--loop expressions for the longitudinal
part of the $W$ boson self--energies in the general case $m_U \neq m_D \neq 0$
given in Ref.~\cite{nous} [eqs.~(2.5,2.8) and (4.1,4.2) of that paper,
respectively] leads to the one and two--loop expressions of the charged Higgs
boson self--energy given in eqs.~(2.6--2.7) and (4.1--4.4) of the present
paper. This is just because for tg$\beta=1$, the charged Higgs boson couples to
fermions exactly like the charged Goldstone of the Standard Model,
up to a relative minus sign for up--type and down--type quarks.  \s

In a completely analogous manner, one can derive the Ward identity in the
case of the neutral Golsdstone boson $\Phi_2$, which writes
\be
q^2 \Pi_{ZZ}^L(q^2)= -M_Z^2 \Pi_{\Phi_2
\Phi_2}(q^2)-\frac{g^2}{4\cos^2\theta_W}
\langle 0| S_1(0)|0\rangle\label{wiw}
\en
which allows to check the expressions of the self--energies of the pseudoscalar
Higgs boson which, again for tg$\beta=1$, has exactly the same couplings as
the neutral Standard Model Goldstone up to, again, a relative minus sign
for isospin up and down quarks. \s

This provides a powerful consistency check of both the calculation of the
electroweak vector boson self--energies performed in Ref.~\cite{nous} and
the one of the neutral pseudoscalar and charged Higgs boson self--energies
presented here.

\vspace*{2cm}

\subsection*{Figure Captions}

\vspace*{0.5cm}

\begin{description}
\item[Fig.~1] Feynman diagrams for the contribution of quark pairs to the
self--energy of a Higgs boson at the one--loop (a) and two--loop (b).

\item[Fig.~2] Feynman diagrams for the one--loop quark self--energy (a) and
for the mass and vertex counterterms  contribution to the self--energies at
the two--loop level (b).

\item[Fig.~3] Tadpole diagrams relating the Higgs boson self--energies to the
vector boson self--energies (a) at one--loop, (b) at two--loop and (c)
mass counterterm contributing at the two--loop level.

\end{description}

\end{document}